\documentclass[11pt]{amsart}

\usepackage{amssymb}
\usepackage{url}
\usepackage{hyperref}

\allowdisplaybreaks[4]

\usepackage{mathrsfs}
\let\mathcal\mathscr
\usepackage[mathscr]{eucal}

\newcommand*{\pd}[2]{\mathchoice{\frac{\partial#1}{\partial#2}}
  {\partial#1/\partial#2}{\partial#1/\partial#2}
  {\partial#1/\partial#2}}
\newcommand*{\od}[2]{\mathchoice{\frac{d#1}{d#2}}
  {d#1/d#2}{d#1/d#2}{d#1/d#2}}
\newcommand*{\fd}[2]{\mathchoice{\frac{\delta#1}{\delta#2}}
  {\delta #1/\delta#2}{\delta#1/\delta#2}{\delta#1/\delta#2}}

\let\phi=\varphi
\let\kappa=\varkappa

\DeclareMathOperator{\const}{const}

\newcommand*{\Ev}{\mathbf{E}}
\newcommand*{\g}{\mathfrak{g}}

\theoremstyle{theorem}

\numberwithin{proposition}{section}

\numberwithin{corollary}{section}

\numberwithin{theorem}{section}

\numberwithin{lemma}{section}
\theoremstyle{definition}

\theoremstyle{remark}

\usepackage{mathrsfs}
\let\mathcal\mathscr
\usepackage[mathscr]{eucal}

\newcommand{\cprime}{\/{\mathsurround=0pt$'$}}

\author{I.S.\,Krasil{\cprime}shchik}
\address{Trapeznikov Institute of Control
    Sciences, 65 Profsoyuznaya street, Moscow 117997,
    Russia}
  \email{josephkra@gmail.com}
\author{V.V.\,Lychagin}
  \address{Trapeznikov Institute of Control Sciences, 65
    Profsoyuznaya street, Moscow 117997,
    Russia}
  \email{lychagin@math.uit.no}
\title[Gas behavior in a one-dimensional nozzle]{Geometric study of gas
  behavior in a one-dimensional nozzle (the case of the van der Waals gas)}
\thanks{Partially supported by Russian Foundation for Basic Research, Grant
   18-29-10013}

\begin{document}

\begin{abstract}
  We construct a three-component system of PDEs describing dynamics of
  van~der~Walls gas in one-dimensional nozzle. The group of conservation laws
  for this system is described. We also compute the Lie algebras of point
  symmetries and present group classification. Examples of exact invariant
  solutions are given.
\end{abstract}

\subjclass[2010]{76N15, 37K05}%
\keywords{Gas dynamics, one-dimensional nozzle, symmetries, conservation
  laws, invariant solution}%
\maketitle
%\setcounter{page}{3}

%\newpage
\setcounter{tocdepth}{3}
%\tableofcontents
%\newpage

\section*{Introduction}
\label{sec:introduction}

Methods of differential geometry applied to the problems arising in analysis
of partial differential equations (PDEs) allow one to construct a lot of
important invariants, such as, e.g., symmetries and conservation laws
(see~\cite{AMS-book}, \cite{KLV}, \cite{Ovs}). In turn, these invariants may
be used for qualitative study of equations, their classification, construction
of exact solutions, etc.

In what follows, we use these methods to study gas behavior in in a one
dimensional nozzle. In Section~\ref{sec:model}, we construct the model and
derive the equations for the case the van~der~Waals
gas. Section~\ref{sec:prel-facts-notat} deals with basic constructions from
the geometry of PDEs necessary for the exposition. Conservation laws of the
obtained system are described in Section~\ref{sec:cosymm-cons-laws}, while
Section~\ref{sec:symmetries} deals with the computation of symmetries and
classification of the equations at hand based on their symmetry
algebras. Finally, we discuss invariant solutions in
Section~\ref{sec:invariant-solutions}.

\section{The model}
\label{sec:model}

We consider a gas flow in a wind tunnel (nozzle) of a variable radius. Assume
that the $x$-axes is the symmetry axis of the tunnel, $A(x)$ is an area of
section at the point $x$ and $(y,z)$ components of the gas flow velocity are
negligibly small in comparison with the $x$-component $u(x,t)$. Then the Euler
equations of the momentum consrevation and the mass conservation laws take the
form
\begin{equation}\label{eq:5}
(A\rho u)_t + (p + A\rho u^2)_x = 0, \quad (A\rho)_t + (A\rho u)_x = 0,
\end{equation}
where $\rho$ is the gas density and $p$ is the pressure. We also assume that
the gas flow is adiabatic, i.e.,
\begin{equation}
  \label{eq:6}
  s_t + us_x = 0,
\end{equation}
where $s(x,t)$ is the specific entropy.

The thermodynamic variables $s$, $v$, $e$, $p$ ,and $T$ satisfy in addition
the state equations
\begin{equation*}
  \label{eq:7}
  p = RT\Pi_v,\quad e = RT^2\Pi_T,\quad s = R(\Pi + T\Pi_T),
\end{equation*}
where $v = \rho^{-1}$ is the specific volume, $e$ being the specific inner
energy, $T$ temperature, $R$ the universal gas constant, while
$\Pi = \Pi(v,T)$ is the Massieu-Plank potential (see, for
example,~\cite{Fortov,Planck}).

Let us introduce potential functions $\phi$ and $\psi$ for
Equations~\eqref{eq:5}:
\begin{gather*}
  \label{eq:8}
  A\rho u = -\phi_x,\qquad p + A\rho u^2 = \phi_t,\\
A\rho  = \psi_x,\qquad A\rho u = -\psi_t.
\end{gather*}
Then System~\eqref{eq:5} takes the form
\begin{gather*}
  \phi_x - \psi_t = 0,\qquad p = \phi_t + \psi_t^2\psi_x^{-1},\\
  \rho = \psi_x A^{-1},\quad u = -\psi_t \psi_x^{-1}.
\end{gather*}
Note the function $\psi(x,t)$ is the first integral of the vector field
$\partial_t + u\partial_x$ and rewrite Equation~\eqref{eq:6} in the form
\begin{equation}\label{eq:10}
  s = S (\psi), 
\end{equation}
for some function $S$.  In order to eliminate the unknown
function~\eqref{eq:10}, we use Equation~\eqref{eq:6} in the form
\begin{equation*}\label{eq:11}
  s_v (v_t + uv_x) + s_T (T_t + uT_x) = 0,
\end{equation*}
assuming that the function $s = s(v,T)$ is given by the state equations.  Then
our system takes the form
\begin{equation*}\label{eq:12}
  \phi_t = p - \phi_x^2\psi_x^{-1},\quad
  \psi_t = \phi_x,\quad
  T_t = -uT_x - \frac{s_v}{s_T}(v_t + uv_x),
\end{equation*}
where
\begin{equation*}\label{eq:13}
  v = A\psi_x^{-1},\qquad
  u = -\phi_x \psi_x^{-1}. 
\end{equation*}

In the case of ideal gas flows we have the following state equations
\begin{equation*}
  p = Rv^{-1}T,\quad
  e = \frac{n}{2}RT,\quad
  s = \ln\left(T^{n/2} v\right) + \const,
\end{equation*}
where $n$ is the degree of freedom.  For the van der Waals gas the state
equations read (see~\cite{LychRoop})
\begin{equation*}
  p = \frac{RTv^2 - a(v - b)}{v^2 (v - b)},\quad
  e = \frac{nRT}{2} - \frac{a}{b},\quad
  s = \ln\left(T^{4n/3} (v - b)^{8/3}\right) + \const.
\end{equation*}
The constant $a$ takes into account the intermolecular forces and the constant
$b$ is the molecular volume, while $n$ is the degree of freedom.

As the result, we arrive to the following system of evolutionary equations
\begin{equation}\label{eq:1}
  \begin{array}{l}
    \phi_t = -\dfrac{\psi_x (aA \psi_x  + ab \psi_x^2 + A^3\phi_x^2 +
               bA^2\phi_x^2 \psi_x + R\tau A^2)}{A^2 (A + b \psi_x)},\\
    \psi_t = -\phi_x,\\[4pt]
    \tau_t = -\dfrac{A(2\tau \psi_x \phi_{xx} + n\phi_x \tau_x \psi_x -
               2\tau\phi_x 
               \psi_{xx}) + 2\tau A_x \phi_x \psi_x  + nb\phi_x 
               \tau_x \psi_x^2}{\psi_x^2 n (A + b \psi_x) },
  \end{array}
\end{equation}
where the temperature~$T$ is relabeled to~$\tau$ and~$\psi$ to $-\psi$.
% \begin{equation*}
%   u = \frac{\phi_x}{\psi_x},\qquad
%   p = \phi_t + \frac{\phi_x^2}{\psi_x}.
% \end{equation*}

We study this system in the forthcoming sections. Note that the case~$a$,
$b=0$ corresponds to the ideal gas.

\section{Preliminary facts and notation}
\label{sec:prel-facts-notat}

We consider the space~$J^\infty(2,3)$ with the coordinates~$x$, $t$, $\phi_{kl}$,
$\psi_{kl}$, $\tau_{kl}$, $k$, $l=0,1,\dots$, where the subscript corresponds
to the partial derivative~$\partial^{k+l}/\partial x^k\partial t^l$, and the
subspace $\mathcal{E}\subset J^\infty(2,3)$ defined by Equations~\eqref{eq:1}
and all their differential consequences. The functions~$x$, $t$, $\phi_k =
\phi_{k0}$, $\psi_k = \psi_{k0}$, and $\tau_k = \tau_{k0}$ may be chosen for
internal coordinates in~$\mathcal{E}$. Denote by~$r_\phi$ and~$r_\tau$ the
right-hand sides of the first and third equations in~\eqref{eq:1},
respectively. Then the total derivatives
\begin{align*}
  D_x&=\pd{}{x} + \sum_{k\geq0}\left(\phi_{k+1}\pd{}{\phi_k} +
       \psi_{k+1}\pd{}{\psi_k} + \tau_{k+1}\pd{}{\tau_k}\right),\\
  D_t&=\pd{}{t} - \sum_{k\geq0}\left(D_x^k(r_\phi)\pd{}{\phi_k} +
       \phi_{k+1}\pd{}{\psi_k} + D_x^k(r_\tau)\pd{}{\tau_k}\right)
\end{align*}
are vector fields on~$\mathcal{E}$. For any function~$Y$ on~$\mathcal{E}$
define its linearization
\begin{equation*}
  \ell_Y(f,g,h) = \sum_{k\geq0} \left(\pd{Y}{\phi_k}D_x(f) +
    \pd{Y}{\psi_k}D_x(g) + \pd{Y}{\tau_k}D_x(h)\right).
\end{equation*}

A symmetry of~$\mathcal{E}$ is a vector field that commutes with~$D_x$
and~$D_t$. Any symmetry is of the form
\begin{equation*}
  \Ev_\mathcal{S} = \sum_{k\geq0}\left(D_x^k(\mathcal{S}_\phi)\pd{}{\phi_k} +
    D_x^k(\mathcal{S}_\psi)\pd{}{\psi_k} +
    D_x^k(\mathcal{S}_\tau)\pd{}{\tau_k}\right), 
\end{equation*}
where the vector-function $\mathcal{S} = (\Phi, \Psi, \Theta)$ must satisfy the
linear system
\begin{equation}
  \label{eq:2}
  D_t(\Phi) = \ell_{r_\phi}(\Phi,\Psi,\Theta),\ D_t(\Psi) =
  -D_x(\Phi),\ D_t(\Theta) = \ell_{r_\phi}(\Phi,\Psi,\Theta).
\end{equation}
We identify symmetries with the corresponding functions~$\mathcal{S}$.

A conservation law of~$\mathcal{E}$ is a differential form~$\omega = X\,dx +
T\,dt$ such that $D_t(X) = D_x(T)$. It is called trivial if there exists a
potential~$Z$ such that~$X = D_x(Z)$, $T = D_t(Z)$. The generating function
of~$\omega$ is the triple $G = (G_\phi,G_\psi,G_\tau)$, where
\begin{equation}
  \label{eq:3}
  G_\phi = \fd{X}{\phi},\quad G_\psi =\fd{X}{\psi}, \quad \mathcal{S}_\tau = \fd{X}{\tau}
\end{equation}
and
\begin{equation*}
  \fd{X}{\phi} = \sum_{k\geq0}(-1)^kD_x^k\left(\pd{X}{\phi_x}\right),
\end{equation*}
etc., are the variational derivatives. To find generating functions, one needs
to solve the system
\begin{align*}
  D_t(G_\phi)&= \sum_{k\geq0}(-1)^{k+1}\left(
               D_x^k\left(\pd{G_\phi}{\phi_k}\right)
               +D_x^k\left(\pd{G_\psi}{\phi_k}\right)
               +D_x^k\left(\pd{G_\tau}{\phi_k}\right)
               \right),\\
  D_t(G_\psi)&= \sum_{k\geq0}(-1)^{k+1}\left(
               D_x^k\left(\pd{G_\phi}{\psi_k}\right)
               +D_x^k\left(\pd{G_\psi}{\psi_k}\right)
               +D_x^k\left(\pd{G_\tau}{\psi_k}\right)
               \right),\\
  D_t(G_\tau)&= \sum_{k\geq0}(-1)^{k+1}\left(
               D_x^k\left(\pd{G_\phi}{\tau_k}\right)
               +D_x^k\left(\pd{G_\psi}{\tau_k}\right)
               +D_x^k\left(\pd{G_\tau}{\tau_k}\right)\right)
\end{align*}
adjoint to~\eqref{eq:2}. A conservaton law is trivial if and only if its
generating function vanishes.

\section{Conservation laws}
\label{sec:cosymm-cons-laws}

Solving this system in the case when~$G$ depends on~$x$, $t$, $\phi_k$,
$\psi_k$, $\tau_k$, $k\leq 2$, we get the solutions
\begin{equation*}
  G^0 = (0, 1, 0),\qquad G^\infty = (0,G_\psi,G_\tau),
\end{equation*}
where\begin{align*} G_\psi=
       &-\frac{1}{n\tau\psi_x^2(b\psi_x+A)^2}\bigg(
         2A^2\tau\Big(\psi_x H - \tau^{\frac{n}{2}}(b\psi_x + A)\frac{\partial
         H}{\partial y}\Big)\psi_{xx}\\
       &+\psi_x^2\Big(2\tau A\frac{\partial A}{\partial
         x} + n\tau_x(b\psi_x+A)^2\Big)H
         -2A\tau\psi_x^3(b\psi_x+A)\frac{\partial H}{\partial\psi}\\
       &-A\tau^{\frac{n}{2}}\psi_x(b\psi_x+A)\Big(2\tau\frac{\partial
         A}{\partial x}
         +n\tau_x(b\psi_x +A)\Big)\frac{\partial H}{\partial y}\bigg),\\
    G_\tau&=\frac{H\psi_x}{\tau}.
\end{align*}
where~$H=H(y,\psi)$ is an arbitrary smooth function and
\begin{equation*}
  y = \frac{b\psi_x + A}{\psi_x}\cdot\tau^{\frac{n}{2}}.
\end{equation*}
Using relations~\eqref{eq:3}, one can reconstruct the corresponding
conservation laws and obtain
\begin{equation*}
  \omega_0 = \psi\,dx + \phi\,dt
\end{equation*}
and
\begin{equation*}
  \omega_H = \frac{2H\ln(y)}{n}\cdot(\psi_x \,dx - \phi_x \,dt).
\end{equation*}

\section{Symmetries}
\label{sec:symmetries}

We compute here point symmetries, i.e., such that the function~$\mathcal{S}$
depends on~$x$, $t$ and~$\phi_k$, $\psi_k$, $\tau_k$ for~$k\leq 2$. Note first
that the second equation in~\eqref{eq:2} means that the
form~$\Psi\,dx - \Phi\,dt$ is a conservation law. Due to the results of
Section~\ref{sec:cosymm-cons-laws}, we have
\begin{equation*}
  \Phi = -\lambda\phi + \frac{2H\ln(y)}{n}\phi_x - D_t(P),\qquad
  \Psi = \lambda\psi + \frac{2H\ln(y)}{n}\psi_x + D_x(P),
\end{equation*}
where~$\lambda = \const$ and~$P=P(x,t,\phi,\psi,\tau,\phi_x,\psi_x,\tau_x)$ is
an arbitrary smooth function. Denote by~$\g$ the Lie algebra of
symmetries. The structure of~$\g$ depends on the values of the parameters that
enter the basic equations\footnote{The full classification includes the
  case~$n=-2$, which is physically senseless and we omit it. So, everywhere
  below~$n\neq-2$}. Note that in all the cases below the algebra~$\g$ contains
the $3$-dimensional Abelian ideal~$\g_0$ spanned by the symmetries
\begin{equation*}
  \mathcal{S}_1 = (1,0,0),\quad \mathcal{S}_2 = (0,1,0),\quad \mathcal{S}_3 = (\phi_t,\psi_t,\tau_t).
\end{equation*}
Describing the algebras, we present only those generators that do not belong
to~$\g_0$. In the description of Lie algebra structures we indicate nonzero
commutators only.

\subsection{$b = 0$, $a \neq 0$}\label{sec:b-=-0}

There are eight subcases.

\subsubsection{$A = p = \const$, $n = 2$}
\label{sec:a-=-p}

The algebra $\g$ is $6$-dimensional. Its generators are
\begin{align*}
  \Phi_4 &= \frac{\psi_x^2 a t+p t (p \phi_x^2+R \tau) \psi_x-\phi p^2}{p^2},
           \quad 
           \Psi_4 = t \phi_x,\\
  \Theta_4 &=
             -\frac{\tau t \phi_x \psi_{xx}}{\psi_x^2} + \frac{\tau t
             \phi_{xx}}{\psi_x} 
             + \frac{t \phi_x \tau_x}{\psi_x}-2 \tau-\frac{2 a \psi_x}{p R};\\ 
  \Phi_5 &= \phi_x,\quad \Psi_5 = \psi_x,\quad \Theta_5 = \tau_x;\\
  \Phi_6 &= x\phi_x-2 \phi,\quad \Psi_6 = x\psi_x\psi, \quad \Theta_6 = x\tau_x
           -2 \tau-\frac{2 a \psi_x}{p R}.
\end{align*}
with the commutators
\begin{equation*}
  [\mathcal{S}_1,\mathcal{S}_4] = -\mathcal{S}_1,\
  [\mathcal{S}_1,\mathcal{S}_6] = -2 \mathcal{S}_1,\
  [\mathcal{S}_2,\mathcal{S}_6] 
  = -\mathcal{S}_2, \
  [\mathcal{S}_3,\mathcal{S}_4] = \mathcal{S}_3, 
  [\mathcal{S}_5,\mathcal{S}_6] = -\mathcal{S}_5.
\end{equation*}

\subsubsection{$A = p = \const$, $n \neq 2$}
\label{sec:a-=-p-1}

We have~$\dim\g = 5$ with the generators
\begin{align*}
  \Phi_4 &= \frac{\psi_x^2 a t+p t (p \phi_x^2+R \tau) \psi_x-p^2 (\phi_x
          x-\phi)}{p^2}, 
          \quad \Psi_4 = t \phi_x-\psi_x x+\psi,\\
        &\quad \Theta_4 =
  \frac{-2 \tau t \phi_x \psi_{xx}+2 \tau t \phi_{xx} \psi_x+(t \phi_x-\psi_x
          x) \tau_x \psi_x n}{\psi_x^2 n};\\ 
  \Phi_5 &= \phi_x, \quad \Psi_5 = \psi_x, \quad \Theta_5 = \tau_x.
\end{align*}
The commutators are
\begin{equation*}
  [\mathcal{S}_2,\mathcal{S}_4] = \mathcal{S}_2,\
  [\mathcal{S}_2,\mathcal{S}_5] = 0, \ 
  [\mathcal{S}_3,\mathcal{S}_4] = \mathcal{S}_3, \
  [\mathcal{S}_4,\mathcal{S}_5] = -\mathcal{S}_5.
\end{equation*}

\subsubsection{$A = pe^{qx}$, $p$, $q \neq 0$, $n = 2$}
\label{sec:a-=-peqx}

One has $\dim\g = 5$. The generators are
\begin{align*} 
  \Phi_4 &=
 \frac{e^{-2qx} \psi_x^2 a t+e^{-qx} \psi_x R p t \tau+\psi_x \phi_x^2
          p^2 t-\phi p^2}{p^2}, \quad \Psi_4 =
          t \phi_x, \\
  &\quad \Theta_4 =
  \frac{\tau t \phi_{xx} \psi_x p R-t \tau \phi_x \psi_{xx} p R+\psi_x (-2
    \psi_x^2 a e^{-qx}+(-2 \psi_x \tau+t \phi_x (q \tau+\tau_x)) p
    R)}{\psi_x^2 p R};\\ 
  \Phi_5 &= \phi_x, \quad \Psi_5 = \psi_x, \quad \Theta_5 = \frac{-2 e^{-qx} q a
          \psi_x-p R (q \tau-\tau_x)}{p R}. 
\end{align*}
The commutators read
\begin{equation*}
  [\mathcal{S}_1,\mathcal{S}_4] = -\mathcal{S}_1, \
  [\mathcal{S}_3,\mathcal{S}_4] = \mathcal{S}_3.
\end{equation*}

\subsubsection{$A = pe^{qx}$, $p$, $q \neq 0$, $n \neq 2$}
\label{sec:a-=-peqx-1}
The algebra of symmetries is $4$-dimensional with the generators
\begin{align*}
  \Phi_4 &=
  \frac{e^{-2qx} \psi_x^2 a q t+e^{-qx} \psi_x R p q t \tau+\psi_x \phi_x^2 p^2 q
          t-\phi p^2 q-\phi_x p^2}{(p^2 q)}, \quad \Psi_4 =
          t \phi_x-\frac{\psi_x}{q}, \\
  &\quad \Theta_4 =
  \frac{2 \tau t \phi_{xx} \psi_x q-2 t \tau \phi_x \psi_{xx} q+(-n (q
    \tau+\tau_x) \psi_x+q t \phi_x (n \tau_x+2 q \tau)) \psi_x}{n \psi_x^2
    q}. 
\end{align*}
and commutators
\begin{equation*}
  [\mathcal{S}_1,\mathcal{S}_4] = -\mathcal{S}_1,\
  [\mathcal{S}_3,\mathcal{S}_4] = \mathcal{S}_3.
\end{equation*}

\subsubsection{$A = (px + r)^q$, $p$, $q \neq 0$, $n = 2$}
\label{sec:a-=-px}

The symmetry algebra is $5$-dimensional and generated by
\begin{align*}
  \Phi_4 &=
  \psi_x^2 (p x+r)^{- q} a t+\psi_x (p x+r)^{-q} R t \tau+\phi_x^2 \psi_x
          t-\phi, \quad \Psi_4 =
          t \phi_x,\\
  &\quad \Theta_4 =
  \frac{-t \tau \phi_x \psi_{xx}}{\psi_x^2} + \frac{t \tau \phi_{xx}}{\psi_x}
    + \frac{t \phi_x \tau_x}{\psi_x} + \frac{q p t \tau \phi_x}{(p x+r)\psi_x}
    - 2 \tau-\frac{2 (p x+r)^{-q} a \psi_x}{R};\\ 
  \Phi_5 &= -2 \phi+\frac{(p x+r)\phi_x}{p}, \quad \Psi_5 = -\psi+\frac{(p x+r)
          \psi_x}{p},\\ 
  &\quad \Theta_5 = \frac{(-2 a p q \psi_x-2 \psi_x) (px+r)^{-q} a p+(\tau_x
    (p x+r)-p \tau (q+2)) R}{p R} 
\end{align*}
with the commutators
\begin{equation*}
  [\mathcal{S}_1,\mathcal{S}_4] = -\mathcal{S}_1,\
  [\mathcal{S}_1,\mathcal{S}_5]  = -2\mathcal{S}_1,\
  [\mathcal{S}_2,\mathcal{S}_5] = -\mathcal{S}_2,\ 
  [\mathcal{S}_3,\mathcal{S}_4] = \mathcal{S}_3.
\end{equation*}

\subsubsection{$A = (px + r)^q$, $p$, $q \neq 0$, $n \neq 2$}
\label{sec:a-=-px-1}

Here $\dim\g = 4$. Generators are
\begin{align*}
  \Phi_4 &= -2 \phi+(q+1) \phi -(q+1) t \psi_x (R (p x+r)^{-q} \tau+\psi_x (p
           x+r)^{-2q}  a+\phi_x^2)\\
         &+\frac{(p x+r) \phi_x}{p},\quad \quad \Psi_4 = -\psi-(q+1) t
           \phi_x+\frac{(p x+r) \psi_x}{p},\\
         &\quad \Theta_4 = \frac{1}{(p x+r) \psi_x^2 n p R}
           \Big(-2 \psi_x^3 a n pq  ((p x+r)^{-q+1}+2  (p x+r) (p x+r)^{-q})\\
         &\qquad -2 R (\tau \psi_x p t (q+1) (p x+r) \phi_{xx}-\tau \phi_x p t
           (q+1) 
           (p x+r) \psi_{xx}\\
         &\quad+(-\frac{1}{2} ((p x+r) \tau_x+p q \tau) (p x+r) n
           \psi_x+(\frac{1}{2}\tau_x n (p x+r)+p q \tau) \phi_x p (q+1) t)
           \psi_x)\Big)
\end{align*}
while the brackets read
\begin{equation*}
  [\mathcal{S}_1,\mathcal{S}_4] = (q-1)\mathcal{S}_1,\
  [\mathcal{S}_2,\mathcal{S}_4] = -\mathcal{S}_2,\ 
  [\mathcal{S}_3,\mathcal{S}_4] = -(q+1)\mathcal{S}_3.
\end{equation*}

\subsubsection{$A$ general, $n = 2$}
\label{sec:a-general-n}

The algebra is $4$-dimensional with the generators
\begin{align*}
  \Phi_4 &=\frac{(\psi_x^2 a t+A t (A \phi_x^2+R \tau) \psi_x-\phi A^2)}{A^2},
          \quad 
          \Psi_4 = t \phi_x,\\
  &\quad \Theta_4 = -\frac{t \tau \phi_x
          \psi_{xx}}{\psi_x^2} + \frac{\tau t \phi_{xx}}{\psi_x} + \frac{t
          \phi_x \tau_x}{\psi_x} + \frac{A_x t \tau \phi_x}{\psi_x A}-2 \tau -
  \frac{2 a \psi_x}{A R}.
\end{align*}
and the commutator
\begin{equation*}
  [\mathcal{S}_3,\mathcal{S}_4] = \mathcal{S}_3.
\end{equation*}

\subsubsection{$A$ general, $n \neq 2$}
\label{sec:a-general-n-1}

Here $\g$ coincides with $\g_0$.

\subsection{$b = 0$, $a = 0$ (the ideal gas)}\label{sec:b-=-0-1}

We have four cases.

\subsubsection{$A = p = \const$}
\label{sec:a-=-p-2}

The $6$-dimensional algebra is generated by
\begin{align*}
  \Phi_4 &= \frac{-\phi p+t \psi_x (\phi_x^2 p+R \tau)}{p}, \quad \Psi_4 = t
          \phi_x, \\
          \quad \Theta_4 &=
  -\frac{2 t \tau \phi_x \psi_{xx}}{\psi_x^2 n} + \frac{2 t \tau
          \phi_{xx}}{\psi_x n} + \frac{t \phi_x \tau_x}{\psi_x}-2 \tau;\\
  \Phi_5 &= \phi_x, \quad \Psi_5 = \psi_x, \quad \Theta_5 = \tau_x;\\
  \Phi_6 &= x\phi_x-2 \phi, \quad \Psi_6 = x\psi_x-\psi, \quad \Theta_6 =
           x\tau_x  -2 \tau. 
\end{align*}

The commutators are
\begin{equation*}
  [\mathcal{S}_1,\mathcal{S}_4] = -\mathcal{S}_1,\
  [\mathcal{S}_1,\mathcal{S}_6] = -2\mathcal{S}_1,\
  [\mathcal{S}_2,\mathcal{S}_6] 
  =  -2\mathcal{S}_2, \
  [\mathcal{S}_3,\mathcal{S}_4] = \mathcal{S}_3, \
  [\mathcal{S}_5,\mathcal{S}_6] =  -\mathcal{S}_5.
\end{equation*}

\subsubsection{$A = pe^{qx}$, $p$, $q \neq 0$}
\label{sec:a-=-peqx-2}

One has $\dim\g = 5$. The algebra is generated by
\begin{align*}
  \Phi_4 &= \frac{\psi_x e^{-qx} R t \tau+p (\psi_x t \phi_x^2-\phi)}{p}, \quad
          \Psi_4 = t \phi_x, \\ \Theta_4 &=
  \frac{2 \tau t \phi_{xx} \psi_x-2 t \tau \phi_x \psi_{xx}+\psi_x (-2 \psi_x
          n \tau+t \phi_x (n \tau_x+2 q \tau))}{n \psi_x^2};\\ 
  \Phi_5 &= \phi_x, \quad \Psi_5 = \psi_x,\quad \Theta_5 = -q \tau+\tau_x.
\end{align*}

The commutators are
\begin{equation*}
  [\mathcal{S}_1,\mathcal{S}_4] = -\mathcal{S}_1, \
  [\mathcal{S}_3,\mathcal{S}_4] = \mathcal{S}_3.
\end{equation*}

\subsubsection{$A = (px + r)^q$, $p$, $q \neq 0$}
\label{sec:a-=-px-2}

One has $\dim\g = 5$ and the algebra is generated by
\begin{align*}
  \Phi_4 &= \psi_x (p x+r)^{-q} R t \tau+\psi_x \phi_x^2 t-\phi, \quad \Psi_4 =
          t \phi_x, \\
  &\quad \Theta_4 =
  -\frac{2 t \phi_x \tau \psi_{xx}}{\psi_x^2 n} + \frac{2 \tau t
    \phi_{xx}}{\psi_x n} 
    +\frac{t \phi_x \tau_x}{\psi_x} + \frac{2 q p t \tau \phi_x}{(p x+r)
    \psi_x n} - 2 \tau;\\
  \Phi_5 &= -2 \phi + \frac{(p x+r) \phi_x}{p}, \quad \Psi_5 = -\psi+\frac{(p
          x+r) \psi_x}{p}, 
          \quad \Theta_5 = \frac{(p x+r) \tau_x-p \tau (q+2)}{p}.
\end{align*}
The commutator relations are
\begin{equation*}
  [\mathcal{S}_1,\mathcal{S}_4] = -\mathcal{S}_1, \
  [\mathcal{S}_1,\mathcal{S}_5]  = -2\mathcal{S}_1,\
  [\mathcal{S}_2,\mathcal{S}_5] = -\mathcal{S}_2,\  
  [\mathcal{S}_3,\mathcal{S}_4] = \mathcal{S}_3.
\end{equation*}

\subsubsection{$A$ general}
\label{sec:a-general}

The $4$-dimensional algebra is generated by
\begin{align*}
  \Phi_4 &= \frac{t \psi_x (\phi_x^2 A+R \tau)-\phi A}{A}, \quad \Psi_4 = t
          \phi_x,\\ 
          &\quad \Theta_4 = 
  -\frac{2 t \tau \phi_x \psi_{xx}}{\psi_x^2 n} + \frac{2 \tau t
          \phi_{xx}}{\psi_x n} + \frac{t \phi_x \tau_x}{\psi_x}
          + \frac{2 t A_x \tau \phi_x}{\psi_x A n} - 2 \tau.
\end{align*}
and the brackets are
\begin{equation*}
  [\mathcal{S}_1,\mathcal{S}_4] = -\mathcal{S}_1,\ 
  [\mathcal{S}_3,\mathcal{S}_4] = \mathcal{S}_3.
\end{equation*}

\subsection{$b \neq 0$}\label{sec:b-neq-0}

The four subcases here are as follows.

\subsubsection{$a = 0$, $A = p = \const$}
\label{sec:a-=-0}

We have $\dim\g = 6$. The algebra is generated by
\begin{align*}
  \Phi_4 &=
  \frac{\phi_x^2 \psi_x^2 b t+(p t \phi_x^2+R t \tau-\phi b) \psi_x-p \phi}{(b
          \psi_x+p}, \quad \Psi_4 = 
          t \phi_x, \\
  &\quad \Theta_4 =
  \frac{2 t p \tau \phi_{xx} \psi_x-2 t p \tau \phi_x \psi_{xx}-(2 (b
    \psi_x+p)) \psi_x (-\frac{1}{2} t \phi_x \tau_x+\psi_x \tau) n}{\psi_x^2 n
    (b \psi_x+p)};\\ 
  \Phi_5 &= \phi_x, \quad \Psi_5 = \psi_x, \quad \Theta_5 = \tau_x;\\ 
  \Phi_6 &= x\phi_x-2 \phi, \quad \Psi_6 = x\psi_x-\psi, \quad \Theta_6 = x\tau_x
          -2 \tau 
\end{align*}
with the commutators
\begin{equation*}
  [\mathcal{S}_1,\mathcal{S}_4] = -\mathcal{S}_1, \
  [\mathcal{S}_1,\mathcal{S}_6] = -2\mathcal{S}_1, \
  [\mathcal{S}_2,\mathcal{S}_6]  = -\mathcal{S}_2, \ 
  [\mathcal{S}_3,\mathcal{S}_4] = \mathcal{S}_3, \
  [\mathcal{S}_5,\mathcal{S}_6] = -\mathcal{S}_5. 
\end{equation*}

\subsubsection{$a = 0$, $A$ general}
\label{sec:a-=-0-1}

The symmetry algebra has $4$ generators
\begin{align*}
  \Phi_4 &=
  \frac{\phi_x^2 \psi_x^2 b t+(A t \phi_x^2+R t \tau-\phi b) \psi_x-A \phi}{b
          \psi_x+A}, \quad \Psi_4 =
          \phi_x t, \\
  &\quad \Theta_4 =
  \frac{1}{\psi_x^2 n (b \psi_x+A)}\Big(2 A \tau t \phi_{xx} \psi_x-2 t A \tau
    \phi_x \psi_{xx}\\
  &+(-2 b n \tau 
    \psi_x^2+(\phi_x \tau_x b n t-2 A n \tau) \psi_x+\phi_x t (A \tau_x n+2
    A_x \tau)) \psi_x\Big) 
\end{align*}
that enjoy the relations
\begin{equation*}
  [\mathcal{S}_1,\mathcal{S}_4] = -\mathcal{S}_1, \
  [\mathcal{S}_3,\mathcal{S}_4] = \mathcal{S}_3.
\end{equation*}

\subsubsection{$a \neq 0$, $A = p = \const$}
\label{sec:a-neq-0}

Dimension of $\g$ is five here. The generators are
\begin{align*}
  \Phi_4 &=
           \frac{\psi_x^3 a b t+p t (b p \phi_x^2+a) \psi_x^2+p^2 (p t
           \phi_x^2+R t 
           \tau-b x \phi_x+b \phi) \psi_x-p^3 (\phi_x x-\phi)}{p^2 (b
           \psi_x+p)},\\  
  \Psi_4 &= t \phi_x-\psi_x x+\psi,\\
  \Theta_4 &= 
             \frac{2 p \tau t \phi_{xx} \psi_x-2 t \phi_x p \tau
             \psi_{xx}-\psi_x \tau_x 
             n (-t \phi_x+\psi_x x) (b \psi_x+p)}{\psi_x^2 n (b \psi_x+p)},\\ 
  \Phi_5 &= \phi_x, \quad \Psi_5 = \psi_x, \quad \Theta_5 = \tau_x.
\end{align*}
They are subject to the relations
\begin{equation*}
  [\mathcal{S}_1,\mathcal{S}_4] = \mathcal{S}_1,\
  [\mathcal{S}_2,\mathcal{S}_4] = \mathcal{S}_2,\
  [\mathcal{S}_3,\mathcal{S}_4] = \mathcal{S}_3, \ 
  [\mathcal{S}_4,\mathcal{S}_5] = -\mathcal{S}_5.
\end{equation*}

\subsubsection{$a \neq 0$, $A$ general}
\label{sec:a-neq-0-1}

One has $\g = \g_0$.

\section{Examples of exact solutions}
\label{sec:invariant-solutions}

Let us now describe two types of exact solutions
\begin{itemize}
\item stationary ones, i.e., independent of $t$,
\item traveling waves in the case $A = \const$.
\end{itemize}

\subsection{Stationary solutions}
\label{sec:stationary-solutions}

Consider the stationary case, i.e., assume $\psi_t = \phi_t = \tau_t = 0$ in
Equations~\eqref{eq:1}. Then the second equation reads~$\phi_x = 0$, i.e.,
$\phi = \const$, the entire system reduces to the sole equation
\begin{equation*}
  \psi_x(aA\psi_x + ab\psi_x^2 + RA^2\tau) = 0.
\end{equation*}
Under the assumption $ab\neq 0$, we obtain three solutions
\begin{equation*}
  \psi_x = -\frac{\big(a + \sqrt{a^2-4abR\tau}\big)A}{2ab},
  \quad  \psi_x = \frac{\big(-a + \sqrt{a^2-4abR\tau}\big)A}{2ab},\quad \psi_x
  = 0 
\end{equation*}
of which only the first one has physical meaning, because by
definition~$\psi_x = -\rho A<0$ (recall that in the final system we relabeled
$\psi$ to $-\psi$). Thus, we obtain
\begin{equation*}
  \rho = \frac{a + \sqrt{a^2 - 4abR\tau}}{2ab},
\end{equation*}
or
\begin{equation*}
  \tau =\frac{a\rho(1 - b\rho)}{R},
\end{equation*}
which gives a simple dependence between temperature and density.

\subsection{Traveling-wave solutions}
\label{sec:trav-wave-solut}

If $A = \const$ then System~\eqref{eq:1} becomes invariant with respect to
$x$-translation symmetry $\mathcal{X} = (\phi_x, \psi_x, \tau_x)$, and one can
consider traveling-wave solutions, i.e., solutions invariant with respect to
$\mathcal{X} + \nu \mathcal{T}$, where $\mathcal{T} = (\phi_t, \psi_t,
\tau_t)$, $\nu$ being the velocity. Such solutions are functions of the
variable~$z = x - \nu t$ and one has
\begin{equation*}
  \pd{}{x} =  \od{}{z},\qquad \pd{}{t} = -\nu\od{}{z}.
\end{equation*}
Consequently,~\eqref{eq:1} reduces to the two equations $ \dot{\phi} = \nu
\dot{\psi}$ and
\begin{equation}
  \label{eq:4}
  \big(\nu^2 A^2b\dot{\psi}^3+(a b + \nu^2A^3)\dot{\psi}^2+(aA-\nu^2 A^2
  b)\dot{\psi}-\nu ^2 A^3+R \tau A^2\big)\dot{\psi} = 0,
\end{equation}
where `dot' denotes the $z$-derivative, while the definitions of the
potentials (see Section~\ref{sec:model}) take the form
\begin{equation*}
  \rho = -\frac{\dot{\psi}}{A},\quad u= \nu,\quad p = -\nu(1 +
  \nu)\dot{\psi}. 
\end{equation*}
Now, solving~\eqref{eq:4} with respect to $\tau$ we obtain dependence of
temperature
\begin{equation*}
  \tau = \frac{(b\rho-1)(\nu^2A^3\rho^2-a\rho-\nu^2A)}{R}
\end{equation*}
on density.

\section*{Acknowledgments}
\label{sec:acknowledgments}

Computations were supported by the \textsc{Jets} software,~\cite{Jets}.

\end{document}